\documentclass[preprint,aps ,nofootinbib]{revtex4}
\usepackage{graphicx}
\pdfoutput=1
\usepackage{epsfig}
\usepackage{amsmath}
\usepackage{amsfonts}
\usepackage{amssymb}
\usepackage{color}
\usepackage{dcolumn}
\usepackage{hyperref}
\usepackage{multirow}

\providecommand{\U}[1]{\protect\rule{.1in}{.1in}}

\newcommand{\f}{\begin{equation}}
\newcommand{\ff}{\end{equation}}
\newcommand{\fa}{\begin{eqnarray}}
\newcommand{\ffa}{\end{eqnarray}}

\begin{document}

\title{Holographic striped superconductor}
\author{Yi Ling $^{1,2,3}$}
\email{lingy@ihep.ac.cn}
\author{Meng-He Wu$^{1,2}$}
\email{mhwu@ihep.ac.cn}
 \affiliation{$^1$Institute of High
Energy Physics, Chinese Academy of Sciences, Beijing 100049,
China\\ $^2$ School of Physics, University of Chinese Academy of
Sciences, Beijing 100049, China\\ $^3$ ~ Shanghai Key Laboratory
of High Temperature Superconductors, Shanghai,200444, China }

\begin{abstract}

We construct a holographic model describing the striped
superconductor (SSC), which is characterized by the presence of
pair density waves (PDW). We explicitly demonstrate that the SSC
phase is implemented as the intertwined phase of charge density
waves (CDW) order and uniform superconducting (SC) order. The
interplay of PDW order, CDW order as well as the uniform SC order
in SSC phase is studied. It is found that the PDW order is
prominent when both CDW order and uniform SC order are balanced.
The critical temperature of CDW becomes higher in the presence of
the uniform SC order, but its charge density amplitude is
suppressed. On the other hand, the SC order is not sensitive to
the presence of CDW order. We also demonstrate that among all the
possible solutions, the black hole in SSC phase has the lowest
free energy and thus is thermodynamically favored.
\end{abstract}
\maketitle
\section{Introduction}
The striped superconductor (SSC) is a special kind of high-$T_c$
superconductors characterized by the presence of pair density
waves (PDW), which is described by a spatially modulated order
parameter\cite{Kudo:2008ivh,Berg:2009ivh,Kudo:2010ivh,Berg:2017ivh,Agterberg:2020ivh}.
The SSC phase plays a key role in understanding the mechanism of
high-temperature superconductivity because in many materials SSC
appears as the intertwined phase of charge density wave (CDW) phase
and the uniform superconducting (SC) phase. The former is due to
the spontaneous breaking of the translational symmetry, while the
latter is due to the spontaneous breaking of $U(1)$ gauge
symmetry. The interplay between CDW phase and SC phase has been
studied in high-$T_c$ superconductors for
decades\cite{Tranquada:1997,Fujita:2002,Hanaguri:2004,Calandra:2011,Denholme:2017,Rahn:2012,comin:2014s,cho:2018cs}.
They exhibit very peculiar and complicated relations which may be
both cooperative and competitive. However, due to the strongly
coupled nature of the system, the theoretical mechanism leading to
these relations maintains mysterious, which prevents us from
understanding the abundant phase structure of high-$T_c$
superconductors. Now, more and more experimental evidences for PDW
have been accumulated such as in cuprate $Bi_2Sr_2CaCu_2O_{8+x}$,
where the PDW order exhibits the same period as
CDW\cite{Hamidian:2016}. Therefore, investigating the formation of
SSC and its features from a theoretical point of view will improve
our understanding on the relationship between CDW phase and SC
phase.

The holographic duality, also known as AdS/CFT correspondence, has
been applied to analyze the fundamental problems in strongly
coupled system in condensed matter physics
\cite{Hartnoll,zaanen2015,ammon2015}. The key point is that a
quantum operator involving dynamics with strong interactions can
be holographically dual to a field in one-dimension higher
spacetime, whose dynamics is well described by the classical
theory of gravity. In particular, recent progress on AdS/CMT has
provided us robust foundation to investigate the relationship
between SC phase and CDW phase. On one hand, various holographic
models have been built for SC since the seminal work in
\cite{Gubser:2008px,Hartnoll:2008vx,Hartnoll:2008kx}, where the
occurrence of superconducting phase transition is signaled by the
formation of the scalar hair in the bulk. On the other hand, the
holographic description of CDW has also been established by
spontaneously breaking the translational
invariance\cite{Donos:2013gda,Donos:2013wia,Withers:2013loa,Ling:2014saa}.

In the early stage of AdS/CMT duality, some references were
motivated to construct holographic models for
SSC\cite{Flauger:2010tv,Erdmenger:2013zaa}. However, in these
models the spatially modulated phase was sourced by a chemical
potential, which means the translational invariance was broken
explicitly instead of spontaneously. Moreover, the full
backreaction to the background was not taken into account. Later,
some efforts have been made to implement the phase diagram with SC
and CDW where the translational symmetry is spontaneously broken
\cite{Kiritsis:2015hoa,Giordano:2018bsf}, and then attempted to
construct a holographic model for PDW
\cite{Cremonini:2016rbd,Cremonini:2017usb}. Perhaps it should be
stressed that in these papers the holographic superconductor is
achieved by means of St\"uckelberg mechanism, rather than the
standard $U(1)$ symmetry breaking in which the scalar field is the
modular of the complex field and thus must be positive definite.
Furthermore, the translational symmetry breaking is characterized
by the same order parameter of superconductivity such that these
two different symmetries must be broken simultaneously
\cite{Cremonini:2016rbd,Cremonini:2017usb}. In another word, the
CDW phase and PDW always coexist and can not be separated such
that SSC as the intertwined phase of CDW and SC is not transparently
demonstrated.

In this paper, we intend to construct a holographic model for SSC
that is achieved as the intertwined phase of CDW phase and SC
phase. In contrast to the previous holographic work in literature,
we insist that the CDW phase is implemented by breaking the
translational invariance spontaneously and the SC phase is
implemented by the standard $U(1)$ symmetry breaking, rather than
St\"uckelberg mechanism. More importantly, we will introduce
different order parameters for above symmetry breaking such that
the CDW and SC phase can exist individually. Moreover, the
interplay between the CDW phase and SC phase can be manifestly
demonstrated such that the SSC as the intertwined phase of these
two phases is manifestly observed in the phase diagram. In another
word, the PDW order is induced due to the coexistence of CDW and
SC. Therefore, the SSC is characterized by the coexistence of
three orders, namely the CDW, the SC and the PDW order, which has
been experimentally observed in Ref.\cite{cho:2018cs}. We
investigate the interplay of these three orders in SSC phase and
explore how the features of PDW could reflect the relationship
between the CDW and the SC.

Nevertheless, the main motivation of the  current paper comes from
our recent work in  Ref.\cite{Ling:2019gjy}, in which we have
successfully constructed a holographic model demonstrating that
the superconductivity can be induced by CDW solely.  To achieve this, one needs to separate  the CDW from free charges such that the
electric chemical potential is set to zero. In this paper we
intend to turn on the electric chemical potential and investigate
the relationship between CDW and SC which is induced by the normal
free charges. Therefore,  in this paper we introduce a doping
parameter $\text{x}$, which is defined as the ratio of two
chemical potentials associated with two gauge fields $A$ and $B$
respectively.  In current paper we will obtain a different kind of
CDW in contrast to the CDW constructed in Ref.\cite{Ling:2019gjy}, and
demonstrate that the CDW exhibits a quite different  relationship
with SC. More explicitly, we find that in the presence of free
charges, only the even orders of the Fourier modes exist in the
expansion of the charge density, while in  Ref.\cite{Ling:2019gjy}
only the odd orders  exist. Based on the results obtained in this
paper and those in Ref.\cite{Ling:2019gjy}, we intend to
conjecture that the complicated relationship between CDW and SC
observed in experiments would be the combinational effects of
these two different kinds of CDWs.

The paper is organized as follows. In Sec.\ref{sec1}, we present
the holographic setup for the striped superconductor based on
doubly charged black holes. In Sec.\ref{seccdw}, we investigate
the instability of the system under the perturbations with
spatially modulated modes, and obtain the phase diagram for CDW
with the critical temperature as the function of the doping
parameter $\text{x}$. In Sec.\ref{secsc}, we investigate the
superconducting condensate in the absence of the translational
symmetry breaking. Then we focus on the SSC phase in
Sec.\ref{seccdwsc}, which results from the coexistence of CDW
phase and SC phase. The full background with charge density waves
and superconductivity will be numerically obtained and the
relations among these three orders will be analyzed. We present
our conclusions and discussions in the last section.

\section{The holographic setup}\label{sec1}
We consider a holographic model in four dimensional spacetime, in
which  gravity is coupled to a dilaton field, two $U(1)$ gauge
fields and a complex scalar field. The action is given by,
\begin{equation}\label{eq:eps+1}
  \begin{aligned}
  S=&\frac{1}{2\kappa^2}\int d^4 x \sqrt{-g} \left[R-\frac{1}{2}\left ( \nabla \Phi  \right )^2-V\left ( \Phi  \right ) -\frac{1}{4}Z_{A}(\Phi )F^2 \right. \\
 & \left. -\frac{1}{4}  G^2 - \left |\left ( \nabla  -ieB  \right )\Psi \right |^2-m_{v}^{2}\Psi \Psi^*\right],
  \end{aligned}
\end{equation}
where $F=dA$, $G=dB$, $Z_{A}(\Phi )=1-\frac{\beta }{2} L^2
\Phi^2$,  $V ( \Phi )=-\frac{1}{L^{2}} + \frac{1}{2}m_{s}^{2}\Phi
^{2}$. The $\beta$-term is introduced to induce the instability of
the homogeneous background such that the translational symmetry
can be spontaneously broken. The dilaton field $\Phi$ is real and
its leading order near the boundary will be treated as the order
parameter of translational symmetry breaking. We also introduce
two $U(1)$ gauge fields $A$ and $B$ such that the black hole could
be doubly charged, but we will treat $B$ as the electromagnetic
field and study the transport properties of its dual field.
Moreover, to break $U(1)$ gauge symmetry spontaneously we
introduce the complex scalar field $\Psi$ as the order parameter
of condensation. We redefine the complex scalar field $\Psi$ as
$\eta e^{i\theta }$, where $\eta>0$ and $\theta$ will be set
$\theta=0$ as a gauge fixing. The equations of motion are given
by,
\begin{equation}\label{eq:eps+31}
  \begin{aligned}
     & R_{\mu \nu }-T_{\mu \nu }^{\Phi }-T_{\mu \nu }^{A }-T_{\mu \nu }^{B }  -T_{\mu \nu }^{\eta  }  =0, \\
     & \nabla  ^{2} \Phi -\frac{1}{4}Z_{A}' F^2-V'=0,                                              \\
     & \nabla  ^{2} \eta  - m_{v}^2 \eta  - (e B)^2\eta  =0,                                                                       \\
     & \nabla_{\mu }(Z_A F^{\mu \nu })=0,                                                                                   \\
     & \nabla_{\mu } G^{\mu \nu }-2 e^2 \eta ^2 B^{\nu } =0,
  \end{aligned}
\end{equation}
where
\begin{equation}\label{eq:eps+4}
  \begin{aligned}
    T_{\mu \nu }^{\Phi }   & =\frac{1}{2} \nabla_{\mu }\Phi \nabla_{\nu  }\Phi +\frac{1}{2}V g_{\mu \nu },                                       \\
    T_{\mu \nu }^{A  }     & =\frac{Z_{A}}{2} \left(F_{\mu \rho  }F^{\rho }_{\nu }- \frac{1}{4} g_{\mu \nu }F^2\right),                          \\
    T_{\mu \nu }^{B  }     & =\frac{1}{2} \left(G_{\mu \rho  }G^{\rho }_{\nu }- \frac{1}{4} g_{\mu \nu }G^2\right),                          \\
    T_{\mu \nu }^{\eta  }  & = \nabla_{\mu }\eta \nabla_{\nu  }\eta  +e^{2}\eta ^{2}B_{\mu }B_{\nu } +\frac{1}{2}m_{v}^{2}\eta ^{2}g_{\mu \nu
    },
  \end{aligned}
\end{equation}
and the prime $'$ denotes the derivative with respect to the
dilaton field $\Phi$. We consider the formation of CDW and
superconductivity over a doubly charged AdS-RN black hole. The
ansatz for the background can be written as:
\begin{equation}\label{eq:eps+5}
  \begin{aligned}
     & ds^{2}=\frac{1}{z^2}\left[-(1-z)p(z)Q_{tt}dt^{2}+\frac{Q_{zz}dz^{2}}{(1-z)p(z)} +Q_{xx}(dx + z^2 Q_{xz}dz)^2 +Q_{yy}dy^{2} \right], \\
     & A_t=\mu_1(1-z)a, \ \  B_t=\mu_2(1-z)b, \ \ \Phi =z\phi,  \ \ \eta =z \zeta   ,
  \end{aligned}
\end{equation}
where $p(z)=4 \left(1+z+z^2-\frac{(\mu_{1}^2+\mu_{2}^2) ^2
z^3}{16} \right)$, $\mu_1$ and $\mu_2$ are the chemical potential
of gauge field $A$ and $B$, respectively. $Q_{tt}$, $Q_{zz}$,
$Q_{xx}$, $Q_{yy}$, $Q_{xz}$, $\psi$, $a$, $b$, $\phi$, $\zeta$
are functions of $x$ and $z$. Obviously, if we set $Q_{tt}$=
$Q_{zz}$= $Q_{xx}$= $Q_{yy}=a=b=1$ and $Q_{xz}=\phi=\zeta=0$, the
background is a doubly charged version of AdS-RN black hole.
Throughout this paper we shall set the AdS radius $l^2=6L^2=1/4$,
the masses of the dilaton and the condensation
$m_{s}^{2}=m_{v}^{2}=-2/l^2=-8$, the coupling constant
$\beta=-129$. In addition, we require that
$\frac{l^2}{2\kappa^2}\gg 1$ such that the large N limit of the
dual field theory is guaranteed. Moreover, we will take $\mu_1$
as the unit and define $\text{x}=\mu_{2}/\mu_{1}$ as the doping
parameter, as proposed in Ref.\cite{Kiritsis:2015hoa}. Obviously,
with a larger doping parameter, the system contains more effective
carriers and its transport properties are expected to vary
correspondingly. Finally, the Hawking temperature of the black
hole is simply given by $T/ \mu_1=(48-\mu_{1}^2-\mu_{2}^2)/(16 \pi
\mu_{1})$ and this expression is also applicable for striped black
holes due to the Einstein-DeTurck
method\cite{Headrick:2009pv,Horowitz:2012ky,Horowitz:2012gs,Horowitz:2013jaa,Ling:2013aya,Ling:2013nxa}.

We remark that the holographic model considered in this paper is
different from what has previously been studied in Ref.
\cite{Ling:2014saa,Ling:2019gjy}, where a coupling term between
gauge field $A$ and $B$, namely $\gamma \Phi FG$, is introduced
into the action such that the dome of the unstable region could
shift to a position with non-zero $k_c$. This coupling term is
harmless there since the gauge field $B$ is turned off, such that
the order parameter of translational symmetry, namely the dilaton
field $\Phi$, always has a zero solution before the translational
symmetry breaking. However, in current paper we intend to
introduce free charges with finite density associated with the
gauge field $B$, thus $B$ is turned on to form a doubly charged
AdS-RN black hole as the background. At this situation, one needs
to turn off the coupling term, otherwise it would contribute a
non-trivial term to the equation of dilaton field $\Phi$ in Eq.
(\ref{eq:eps+31}), such that the order parameter would not have a
zero solution even prior to the symmetry breaking. Therefore, in
current paper we drop off this coupling term and demonstrate that
it will lead to interesting phenomenon for the interplay between
CDW and SC, which is dramatically different from those observed in
\cite{Ling:2014saa,Ling:2019gjy}.

Near the AdS boundary $z\rightarrow 0$, we obtain the following
asymptotic form of the fields,
\begin{equation}\label{eq:eps+3}
  \begin{aligned}
    & Q_{tt}=1+q_{tt}(x)z^{3}+o(z^4), \ Q_{zz}=1+o(z^4), \\
 & Q_{xx}=1+q_{xx}(x)z^{3}+o(z^4), \ Q_{yy}=1+q_{yy}(x)z^{3}+o(z^4), \\
 & Q_{xz}(z)=o(z^2), \ A_t=\mu_1-\rho_A(x)z +o(z^2), \\
& B_t=\mu_2-\rho_B(x)z +o(z^2),
  \end{aligned}
\end{equation}
where $\rho_B(x)$ is the charge density we are interested in, and
the form of Fourier expansion is given by
\begin{eqnarray}\label{eq:eps+6}
\rho_B(x)=\rho^{(0)}_{B}+\rho^{(1)}_{B} \cos (k x)+\rho^{(2)}_{B}\cos (2 k x)+\cdots.
 \end{eqnarray}
In Sec. \ref{seccdw}, \ref{secsc} and \ref{seccdwsc}, we will
firstly obtain the phase diagram for the system by perturbative
analysis and then solve all the coupled equations with full
backreaction numerically. Before this we argue that over a doubly
charged black hole background, both translational symmetry and
$U(1)$ gauge symmetry could be spontaneously broken when the
temperature drops down, giving rise to the CDW phase and SC phase,
respectively. We will explicitly show that, which one breaks prior
to the other depends on the doping parameter $\text{x}$.

\section{The charge density wave phase ($\Phi \neq 0$,\ \ $\eta=0$ )} \label{seccdw}

\begin{figure} [h]
  \center{
  \includegraphics[scale=0.6]{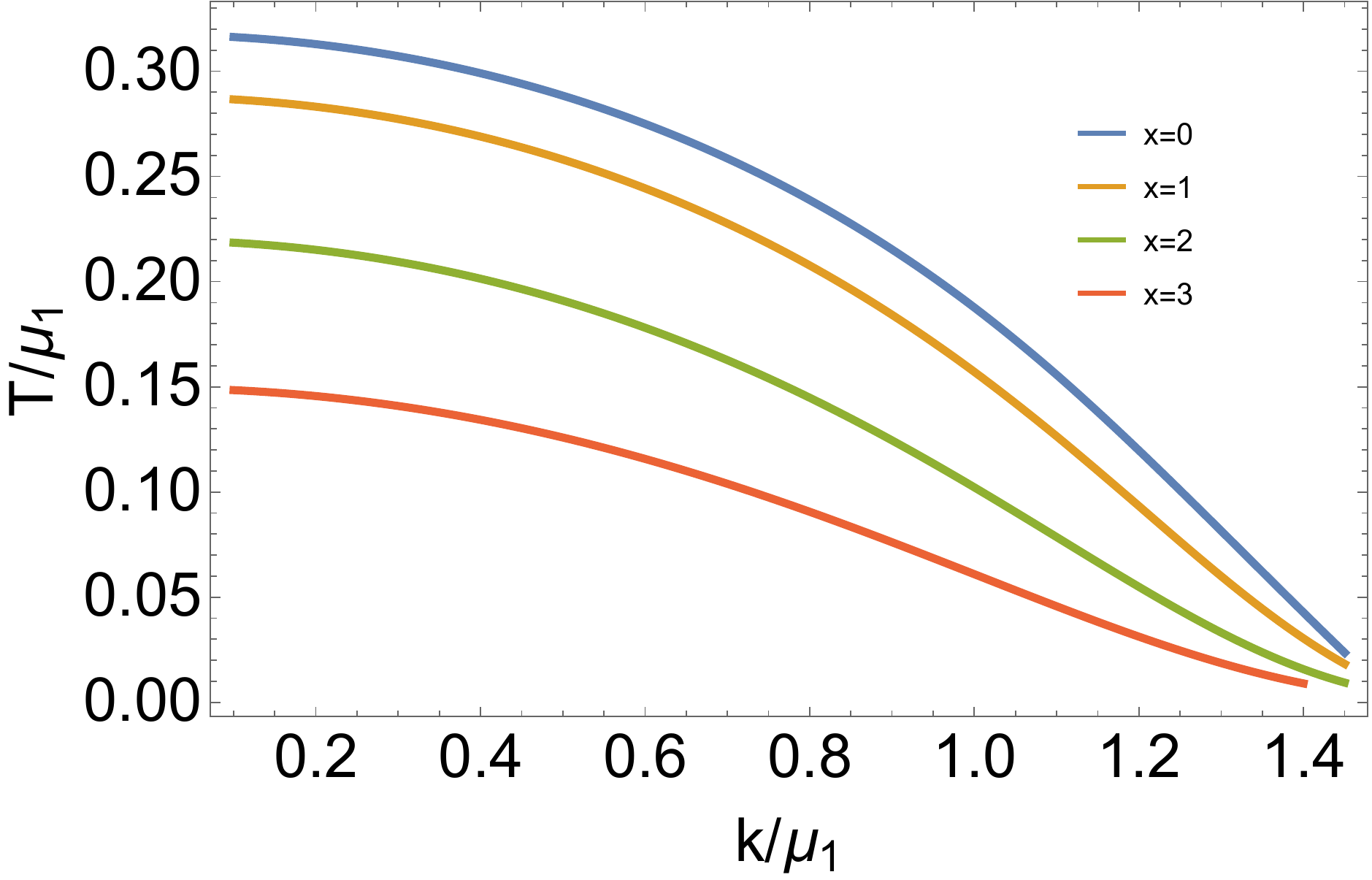}\ \hspace{0.05cm}
  \caption{\label{fig1} The critical temperature as the function of wave number for $\text{x}=0,1,2,3$. The region below each curve becomes unstable and CDW can be generated. }}
\end{figure}

In this section we consider the formation of CDW in the absence of
superconductivity, which can be done by turning off the condensate
field, namely setting $\eta=0$ always. The holographic CDW is formed
by spontaneously breaking the translational symmetry. Such
instability is caused by the violation of the BF bound of $AdS_2$,
which is the near horizon geometry in the extremal limit. Below
the critical temperature $T_c$, the AdS-RN black hole becomes
unstable and the spatially modulated modes may appear to form a
striped black hole. Specifically, we turn on the following linear
perturbation to examine the instability of the black
hole,\footnote{Unlike the case in Ref.\cite{Ling:2019gjy}, here
one need not turn on the perturbation of gauge field $B$ since at
linear level they are decoupled to each other.}
\begin{eqnarray}\label{eq:eps+7}
\delta \Phi  =  \delta \phi (z) \cos(k_cx).
\end{eqnarray}
By substituting (\ref{eq:eps+7}) into the equations of motion
(\ref{eq:eps+31}), we obtain the linearized equation for $\delta \Phi
$. Moreover, we demand the regular boundary condition at the
horizon $z=1$, while the asymptotic expansion of $\delta \phi (z)$
near the boundary is
\begin{eqnarray}\label{eq:eps+8}
\delta \phi \approx \delta \phi _{1} z + \delta \phi _{2} z^{2}+ \cdots,
\end{eqnarray}
where $\delta \phi _{1}$ is treated as the source and  $\delta
\phi _{2}$ as the expectation value in the dual theory. Since one
expects the translational invariance is spontaneously breaking, we
set $\delta \phi _{1}=0$. In Fig.\ref{fig1}, we plot the critical
temperature as the function of the wave number for
$\text{x}=0,1,2,3$. Below the curves the AdS-RN black hole
becomes unstable and the spatially modulated modes will appear
with the allowed wave number as illustrated in Fig.\ref{fig1}.

\begin{figure} [t]
  \center{
  \includegraphics[scale=0.5]{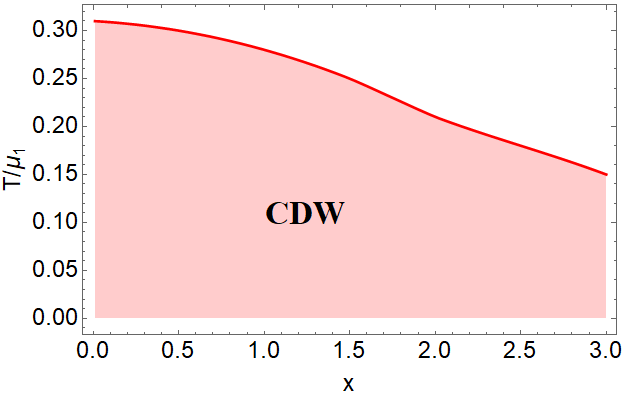}\ \hspace{0.05cm}
  \caption{\label{fig2} The CDW phase diagram in $(\text{x},T)$ plane.}}
\end{figure}

Without loss of generality, we take the wave number
$k_c/\mu_1=0.34$ throughout the paper. For a given $k_c/\mu_1$ and
doping parameter $\text{x}$, one can drop down the temperature of
the black hole, then the instability of the background is signaled
by the presence of non-trivial solution for $\delta \Phi$ at some
temperature, which is estimated as the critical temperature $T_c$.
As illustrated in Fig.\ref{fig2}, the curve in red depicts the
critical temperature $T_c$ of CDW phase as the function of
$\text{x}$. In general, we find the critical temperature goes down
with the increase of the doping parameter $\text{x}$. This is not
surprising since the CDW is an insulating phase, with the increase
of the carriers, the formation of CDW phase becomes harder.

 \begin{figure} [t]
  \center{
   \includegraphics[scale=0.33]{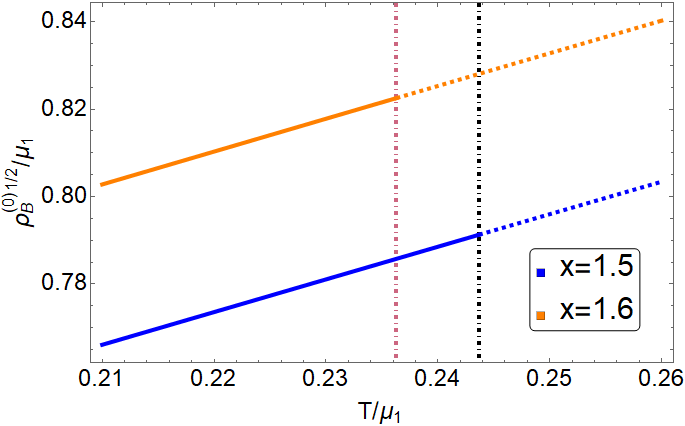}\ \hspace{0.05cm}
   \includegraphics[scale=0.33]{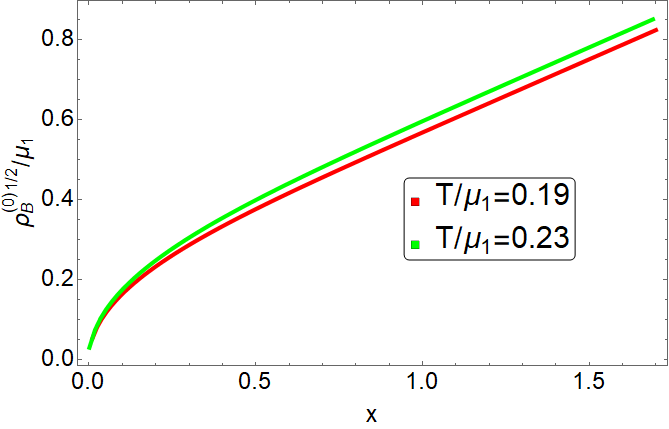}\ \hspace{0.05cm}
  \caption{\label{fig3} The charge
density $\rho^{(0)}_{B}$ as the function of temperature $T$(left)
and doping parameter $\text{x}$(right). The solid lines in
color stand for $\rho^{(0)}_{B}$ in black hole with CDW, while the
dashed lines in color stand for $\rho^{(0)}_{B}$ in RN black
hole. The vertical lines denote the location of CDW phase
transition.}}
\end{figure}

 \begin{figure} [t]
  \center{
  \includegraphics[scale=0.35]{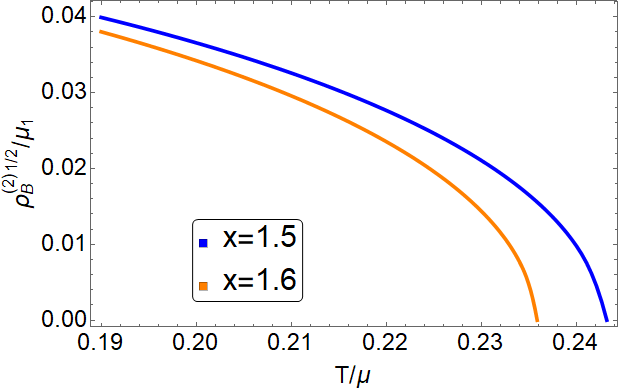}\ \hspace{0.05cm}
  \includegraphics[scale=0.33]{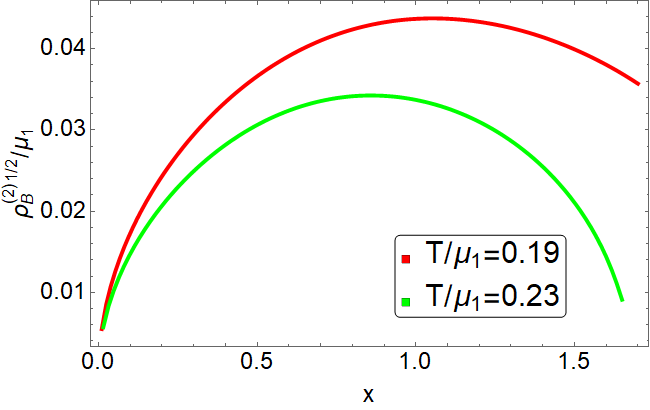}\ \hspace{0.05cm}
  \caption{\label{fig15} The CDW amplitude $\rho^{(2)}_{B}$ as the functions of temperature $T$(left) and doping parameter $\text{x}$(right). }}
\end{figure}

Next, we explicitly construct the background with CDW by
numerically solving all the equations of motion in the absence of
condensation field. In particular we obtain the numerical results
for the charge density. It is found that only {\it even} orders of
the Fourier series of $\rho_B(x)$ appear, namely
\begin{eqnarray}\label{eq:eps+30}
\rho_B(x)=\rho^{(0)}_{B} +\rho^{(2)}_{B}\cos (2 k_c x)+\cdots.
 \end{eqnarray}

It is quite interesting to compare the above result with that
obtained in \cite{Ling:2019gjy}, where only {\it odd} orders of
the Fourier series exist, namely
\begin{eqnarray}\label{eq:eps+110}
\rho_B(x)=\rho^{(1)}_{B}\cos (k_c x)+\rho^{(3)}_{B}\cos (3 k_c x)+\cdots.
 \end{eqnarray}
This key difference results from the following two facts. First,
in \cite{Ling:2019gjy} the chemical potential is always set to
zero, namely $\text{x}=0$, then the leading term of charge density has to
be zero, namely $\rho^{(0)}_{B}=0$. Second, the coupling term $\Phi FG$ is
absent in current paper, thus even in the limit $\text{x}\rightarrow 0$,
the striped black hole background will not go back to the
solutions in \cite{Ling:2019gjy}, instead the charge density
maintains the form  as in Eq.(\ref{eq:eps+30}) and goes to zero in the
limit $\text{x}\rightarrow 0$.

To justify this we may plot the constant term of the charge
density $\rho^{(0)}_{B}$ as well as the CDW amplitude
$\rho^{(2)}_{B}$ as the function of temperature $T$ and doping
parameter $\text{x}$, as illustrated in Fig.\ref{fig3} and
Fig.\ref{fig15}. From the left of Fig.\ref{fig3}, it is obvious to
see that $\rho^{(0)}_{B}$ has the same tendency with the
temperature even in the presence of CDW. In another word, the
presence of CDW does not change the value of $\rho^{(0)}_{B}$ at a
given temperature. Moreover, with the increase of doping
parameter, $\rho^{(0)}_{B}$ increases as well, as illustrated in
the right plot of Fig.\ref{fig3}. In the left plot of
Fig.\ref{fig15}, we find the amplitude of CDW $\rho^{(2)}_{B}$
increases and then intends to be saturated when dropping down the
temperature. In the right plot of Fig.\ref{fig15},
$\rho^{(2)}_{B}$ exhibits interesting behavior with the doping
parameter $\text{x}$. In the limit $\text{x}\rightarrow 0$, both
$\rho^{(0)}_{B}$ and $\rho^{(2)}_{B}$ are vanishing such that the
gauge field $B$ has the zero solution only, while for large
$\text{x}$ we find $\rho^{(2)}_{B}$ becomes smaller again because
the phase transition is suppressed by the increase of carriers, as
illustrated in the phase diagram Fig.\ref{fig2}.

 \begin{figure} [t]
  \center{
   \includegraphics[scale=0.38]{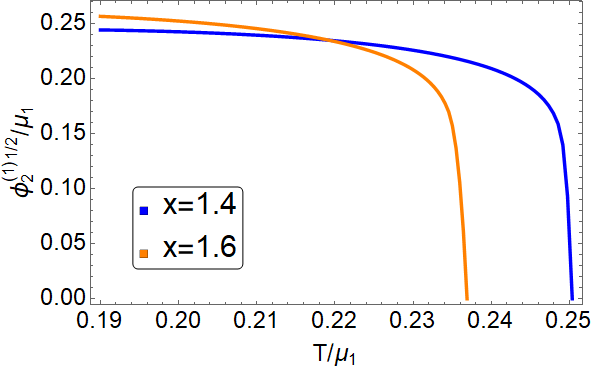}\ \hspace{0.05cm}
   \includegraphics[scale=0.31]{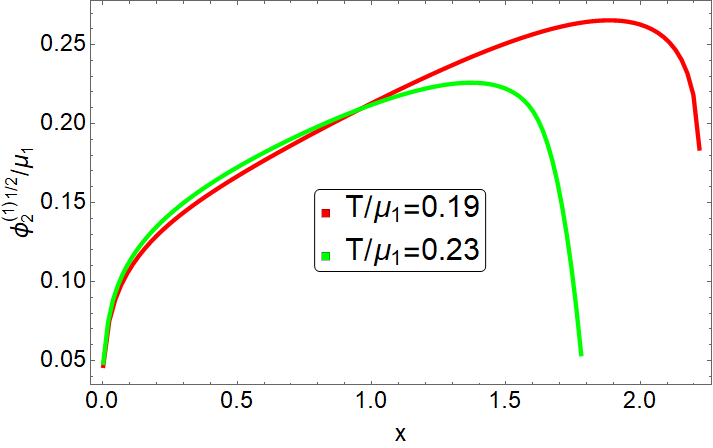}\ \hspace{0.05cm}
  \caption{\label{fig16} The order parameter $\phi _{2}^{(1)}$ as the function of temperature $T$(left) and doping parameter $\text{x}$(right). }}
\end{figure}

The near boundary expansion of the CDW order parameter is $\phi
=\phi _{2}(x) z^2$, and we numerically find $\phi _{2}$ behaves as
\begin{eqnarray}\label{eq:eps+11}
\phi _{2}(x)=\phi _{2}^{(1)}\cos (k_{c}x)+\phi _{2}^{(3)}\cos
(3k_{c}x) \cdots .
 \end{eqnarray}
We plot the magnitude of the order parameter $\phi _{2}^{(1)}$ in
Fig.\ref{fig16}. It is also noticed that $\phi _{2}^{(1)}$ goes to
zero in the limit $\text{x}\rightarrow 0$, while for large
$\text{x}$, $\phi _{2}^{(1)}$ drops down again which behaves
similarly as $\rho^{(2)}_{B}$.

As a summary of this section, we find that the critical
temperature $T_c$ of CDW decreases with the increase of the doping
parameter $\text{x}$, which means achieving such an insulating phase
becomes harder with the doping. However, the Fourier series of the
charge density $\rho_B(x)$ grow up with $\text{x}$, indicating that the
CDW phase benefits from the increase of carriers.

\section{The uniform superconducting  phase ($\eta \neq 0$,\ \ $\Phi=0$ )}\label{secsc}

In this section we consider the condensate of superconductivity in
the absence of CDW, which can be done by setting $\Phi=0$ always.
The SC phase is characterized by the condensation of the complex
scalar fields when $U(1)$ symmetry is spontaneously broken. In
order to evaluate the critical temperature for the condensation,
we may consider the perturbation of the scalar field $\eta$ over a
fixed background. The perturbative equation of motion is given by:
\begin{eqnarray}\label{eq:eps+9}
-\nabla  ^{2} \eta  + m_{v}^2 \eta   = - e^2 B^2\eta.
\end{eqnarray}
As investigated in \cite{Horowitz:2013jaa,Ling:2014laa}, this is a
positive self-adjoint eigenvalue problem for $e^2$. We demand
regularity on the horizon $z=1$. And the asymptotical behavior
near $z=0$ is given by
\begin{eqnarray}\label{eq:eps+10}
 \eta \approx \eta _{1} z + \eta _{2} z^{2}+ \cdots.
\end{eqnarray}
Since $U(1)$ symmetry should be broken spontaneously, we set the
source term $\eta _{1}=0$. The condensation of the scalar field
will be signaled by the appearance of non-vanishing solution for
$\eta _{2}$. For a fixed $e^2$, the corresponding critical
temperature $T_c$ for SC depends on the doping parameter
$\text{x}$.
\begin{figure} [h]
  \center{
  \includegraphics[scale=0.45]{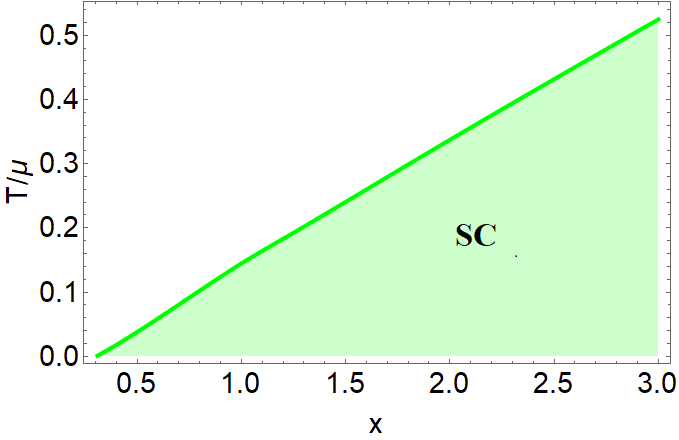}\ \hspace{0.05cm}
  \caption{\label{fig4} The SC phase diagram in $(\text{x},T)$ plane.}}
\end{figure}
As illustrated in Fig.\ref{fig4}, the curve in green depicts the
critical temperature $T_c$ of SC phase as the function of
$\text{x}$ for $e=4$. In contrast to CDW phase, we find that with
the increase of the doping parameter $\text{x}$, the critical
temperature $T_c$ for SC phase goes up, indicating that the
condensation of Cooper pairs becomes easier with the increase of
carriers.
\begin{figure} [h]
  \center{
   \includegraphics[scale=0.35]{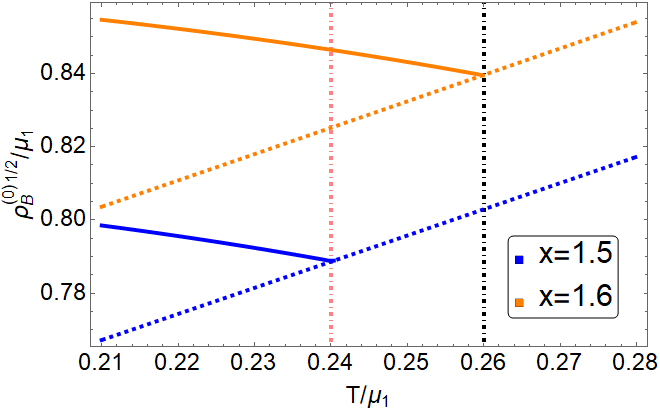}\ \hspace{0.05cm}
  \includegraphics[scale=0.3]{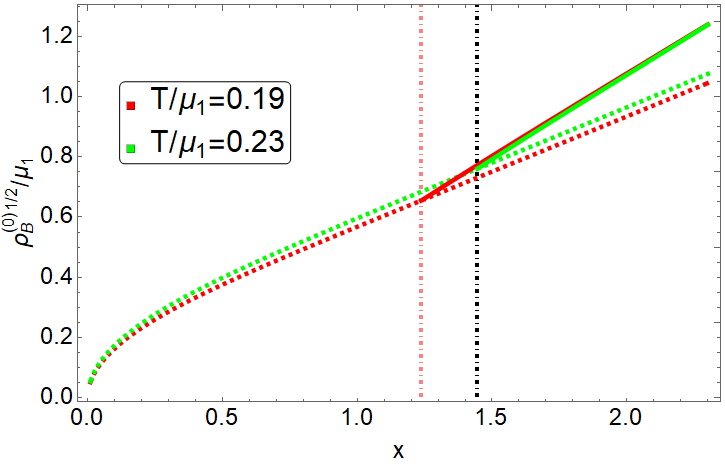}\ \hspace{0.05cm}
\caption{\label{fig5}The charge density $\rho^{(0)}_{B}$ as the
function of temperature $T$ (left) and doping parameter
$\text{x}$(right). The solid lines in color stand for
$\rho^{(0)}_{B}$ after condensation, while the dashed lines in
color stand for $\rho^{(0)}_{B}$ in RN black hole. The vertical
lines denote the location of SC phase transition.}}
\end{figure}

\begin{figure} [h]
  \center{
  \includegraphics[scale=0.36]{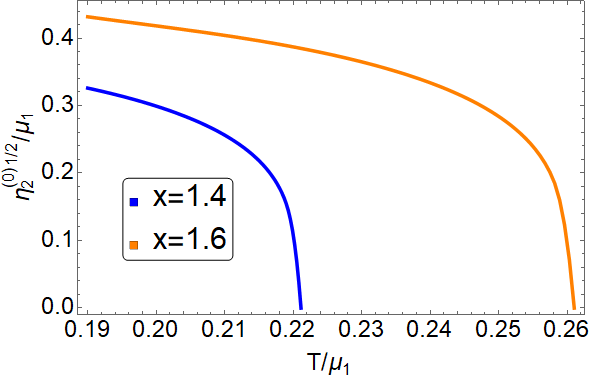}\ \hspace{0.05cm}
  \includegraphics[scale=0.36]{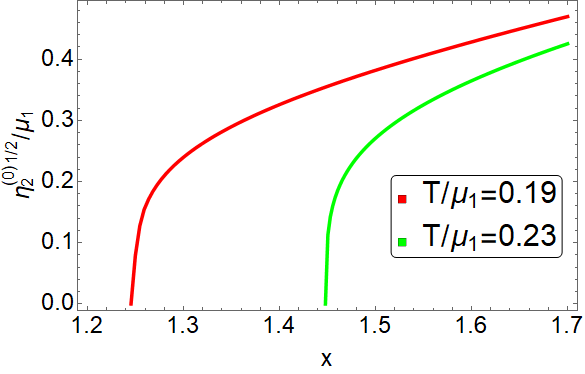}\ \hspace{0.05cm}
  \caption{\label{fig17} The order parameter $\eta _{2}^{(0)}$ of uniform SC as the function of temperature $T$(left) and doping parameter $\text{x}$(right).
  }}
\end{figure}

Next we construct the background with condensation by solving all
the equations of motions. Note that, as compared to
Sec.\ref{seccdw}, the order parameter $\eta _{2}(x)$ of SC and the
charge density $\rho_B(x)$ are now uniform in x-direction, namely
$\eta _{2}(x)=\eta _{2}^{(0)}$ and $\rho_B(x)=\rho^{(0)}_{B}$. We plot
$\rho^{(0)}_{B}$ and $\eta _{2}^{(0)}$ as the function of $T$ and x in
Fig.\ref{fig5} and Fig.\ref{fig17}. Obviously, one finds that below the critical
temperature, the charge density $\rho^{(0)}_{B}$ in SC phase exhibits an
opposite behavior with the temperature and is much larger than
that in normal phase. In addition, we find the charge density as
well as the critical temperature of SC increases with the doping
parameter $\text{x}$. In the plot of Fig.\ref{fig17}, we find
the saturated value of the condensation increases as well with the
doping parameter.

As a summary of this section we conclude that the condensation of
uniform SC benefits from the doping mechanism. The critical
temperature $T_c$ increases with the doping parameter $\text{x}$.
The charge density $\rho^{(0)}_{B}$ and the saturated value $\eta
_{2}^{(0)}$ of the condensation grow up with $\text{x}$ as
well.

\section{The striped SC phase ($\Phi \neq 0$,\ \ $\eta \neq 0$ )}\label{seccdwsc}

\subsection{The phase diagram}\label{sscpd}
\begin{figure} [h]
  \center{
  \includegraphics[scale=0.7]{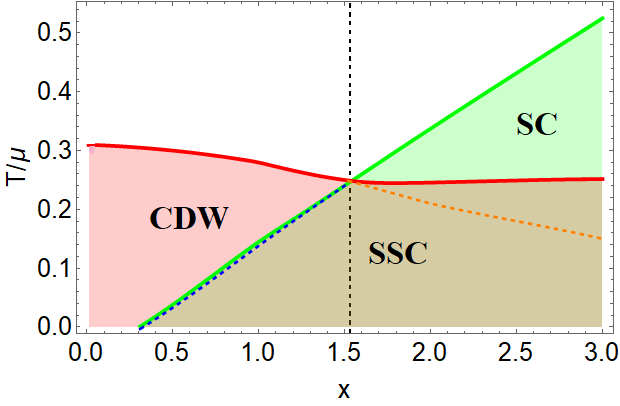}\ \hspace{0.05cm}
  \caption{\label{fig6} The  phase diagram in $(\text{x},T)$ plane. The dashed vertical line
corresponds  to $\text{x}_c\approx 1.53$.}}
\end{figure}
In this section we focus on the striped superconductivity due to
the coexistence of CDW and uniform SC, namely $\eta\neq0$ and
$\Phi \neq 0$. When both the transitional symmetry and $U(1)$
symmetry are spontaneously broken, a new phase is formed which is
called the striped superconducting phase. Now the asymptotic
behavior of $\eta_2$ becomes $x$-dependent and can be expanded as
\begin{eqnarray}\label{eq:eps+11}
\eta_2(x)=\eta_2^{(0)}+\eta_2^{(1)} \cos( k_c x)+\eta_2^{(2)}
\cos(2 k_c x)+\cdots,
 \end{eqnarray}
where among coefficients $\eta_2^{(i)}(i=1,2,...)$ of the
modulated modes, the leading orders, for instance $\eta_2^{(1)}$
or $\eta_2^{(2)}$, could be treated as the order parameter of the
striped SC phase. Obviously, the presence of the striped phase
results from the coexistence of the CDW phase and the SC phase.
Usually this novel phase is also called pair density waves (PDW)
phase implying the periodic distribution of Cooper pairs in
spatial directions. In a parallel way, we may plot the phase
diagram by perturbative analysis, as illustrated in Fig.\ref{fig6}.
The brown region stands for SSC phase, and there are three orders
in this region, namely CDW order, uniform SC order and PDW order,
where $\text{x}_c\approx 1.53$ is the critical doping parameter.
The dashed curve in orange depicts the critical temperature $T_c$
of CDW phase in the absence of SC phase, while the curve in red
depicts $T_c$ of CDW in the presence of SC phase. On the left-hand
side of $\text{x}_c$, the CDW phase is formed prior to the SC
phase, thus these two curves are overlapped, while on the
right-hand side of $\text{x}_c$, the SC phase is formed prior to
the CDW phase, thus we find these two curves are different. In
particular, we find the critical temperature of CDW in the
presence of SC is always higher than that in the absence of
superconductivity. This is because the constant term of the charge
density becomes larger in the presence of SC such that the
formation of CDW becomes easier. Furthermore, the curve in green
depicts the critical temperature $T_c$ of SC phase in the absence
of CDW phase, while the dashed curve in blue depicts $T_c$ of SC
in the presence of CDW phase. It is quite interesting to notice
that on the left-hand side of $\text{x}_c$, the green curve is
almost overlapped with the dashed blue curve, indicating that the
presence of CDW does not change the critical temperature of SC.
This result will further be justified by our analysis on the
background with full backreactions, as presented below.

\begin{figure} [h]
  \center{
  \includegraphics[scale=0.35]{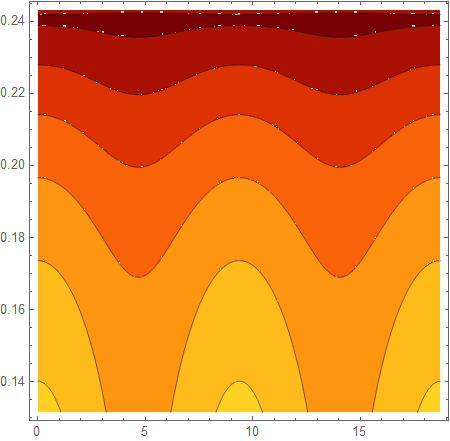}\ \hspace{0.05cm}
  \includegraphics[scale=0.35]{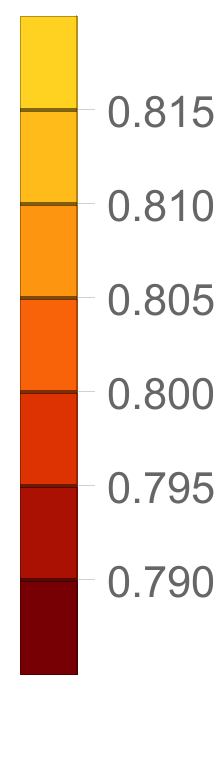}\ \hspace{0.05cm}
   \includegraphics[scale=0.35]{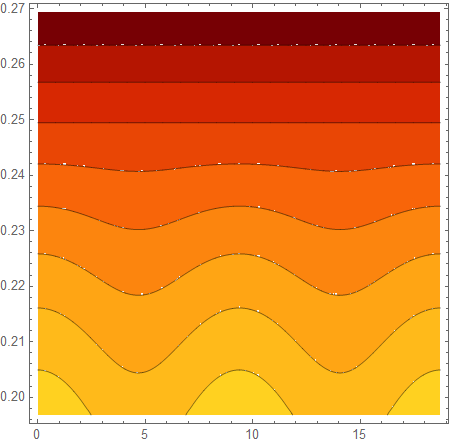}\ \hspace{0.05cm}
  \includegraphics[scale=0.35]{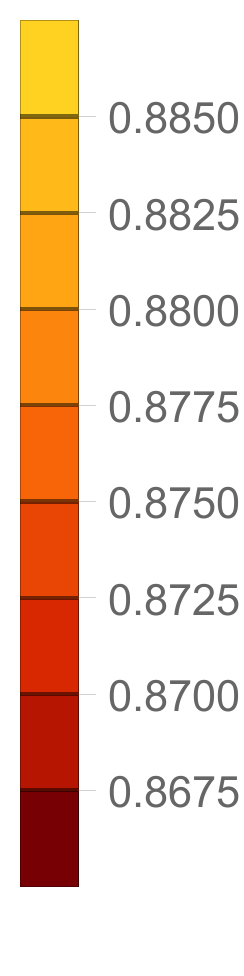}\ \hspace{0.05cm}
  \caption{\label{fig7} The contour plot for the charge density wave $\rho_B(x)$ as a function of $x$ and temperature $T$ for $\text{x}=1.5$ and $\text{x}=1.6$, respectively.}}
\end{figure}

\begin{figure} [h]
  \center{
  \includegraphics[scale=0.35]{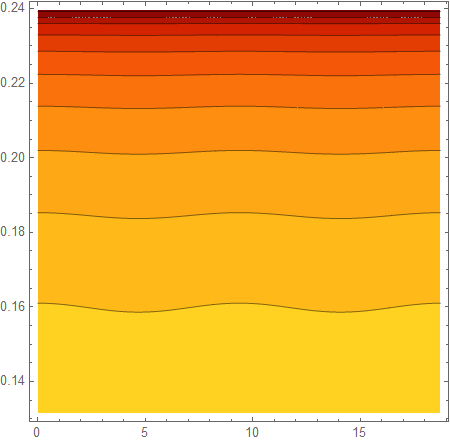}\ \hspace{0.05cm}
  \includegraphics[scale=0.35]{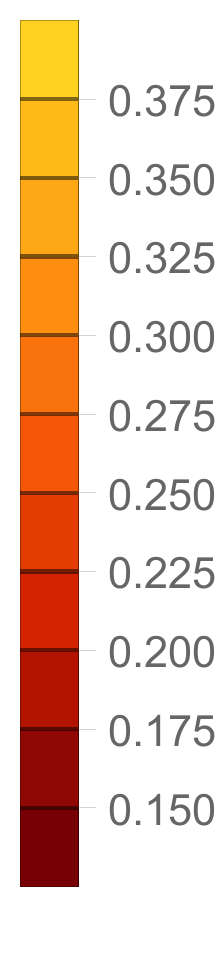}\ \hspace{0.05cm}
   \includegraphics[scale=0.32]{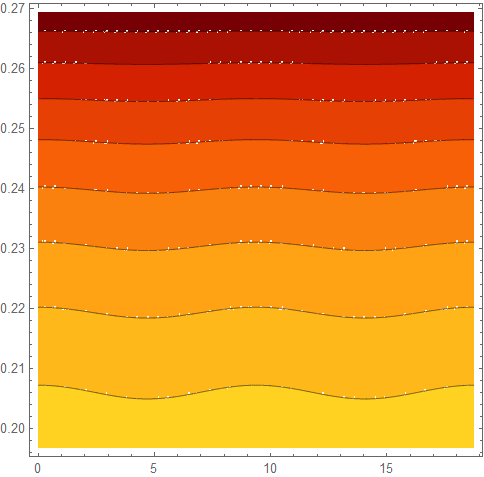}\ \hspace{0.05cm}
  \includegraphics[scale=0.32]{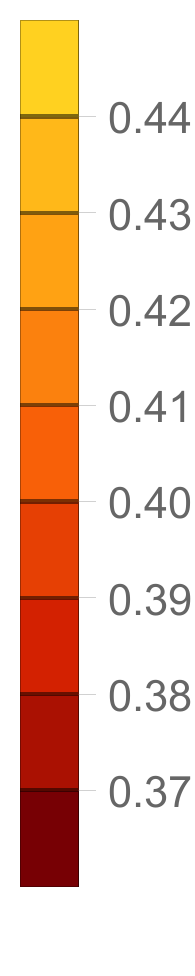}\ \hspace{0.05cm}
  \caption{\label{fig8} The contour plot for  the  the condensate $\eta_2(x)$ as a function of $x$ and temperature $T$ for $\text{x}=1.5$ and $\text{x}=1.6$, respectively. }}
\end{figure}

\subsection{The PDW order in SSC phase }\label{pdwssc}
Now we take the backreactions into account and numerically solve
all the equations for the background with CDW and SC, focusing on
the temperature behavior of the charge density wave $\rho_B(x)$,
the scalar condensate $\eta_2(x)$ and  the CDW order parameter
$\phi _{2}(x)$.

In Fig.\ref{fig7}, we perform the contour plot for the charge
density wave $\rho_B(x)$ as the function of $x$ and temperature
$T$ for $\text{x}=1.5$ and $\text{x}=1.6$, respectively. In
parallel, we perform the contour plot for the condensate
$\eta_2(x)$ as the function of $x$ and temperature $T$ for
$\text{x}=1.5$ and $\text{x}=1.6$ in Fig.\ref{fig8}, respectively.
One finds that $\eta_2(x)$ becomes spatially modulated when the
CDW is involved, indicating that the striped SC is formed.
Numerically, in terms of Fourier series, we find that $\rho_B(x)$,
$\eta_2(x)$, and $\phi _{2}(x)$ have the following form
\begin{eqnarray}\label{eq:eps+11}
\rho_B(x)&=&\rho^{(0)}_{B} +\rho^{(2)}_{B}\cos (2 k_c x)+\cdots,\nonumber\\
 \eta_2(x)&=&\eta_2^{(0)} +\eta_2^{(2)} \cos(2
k_c x)+\cdots,\nonumber\\
\phi _{2}(x)&=&\phi _{2}^{(1)}\cos (k_{c}x)+\phi _{2}^{(3)}\cos
(3k_{c}x) \cdots .
 \end{eqnarray}
The formation of SSC is signaled by the appearance of non-zero
$\eta_2^{(2)}$, that we treat as the order parameter of PDW.
Obviously, the formation of PDW results from the coexistence of
CDW $\phi _{2}^{(1)}$ and the uniform SC $\eta_2^{(0)}$, thus the
SSC is implemented as the intertwined phase of CDW and the uniform
SC phase and is characterized by the coexistence of three orders,
namely CDW order, uniform SC order and PDW order. Moreover, it is
interesting to notice that the condensate $\eta_2(x)$ shares the
same period with the charge density wave $\rho_B(x)$.

\begin{figure} [h]
  \center{
  \includegraphics[scale=0.385]{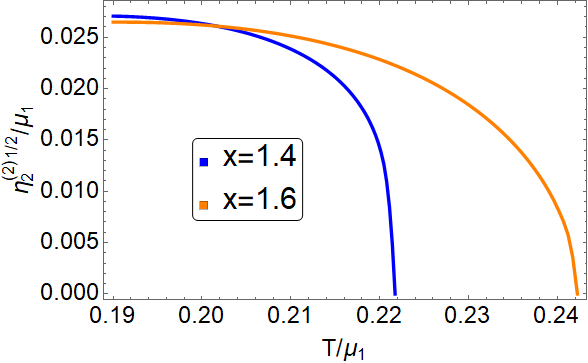}\ \hspace{0.05cm}
  \includegraphics[scale=0.34]{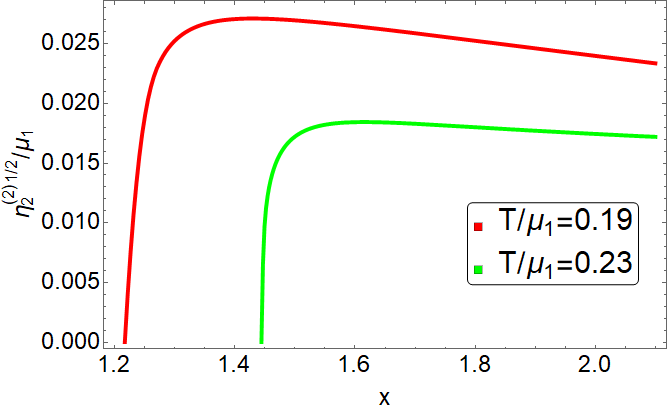}\ \hspace{0.05cm}
  \caption{\label{fig9} The PDW order parameter $\eta_2^{(2)}$ as the function of temperature $T$(left) and doping parameter $\text{x}$(right). }}
\end{figure}

\begin{figure} [h]
  \center{
  \includegraphics[scale=0.35]{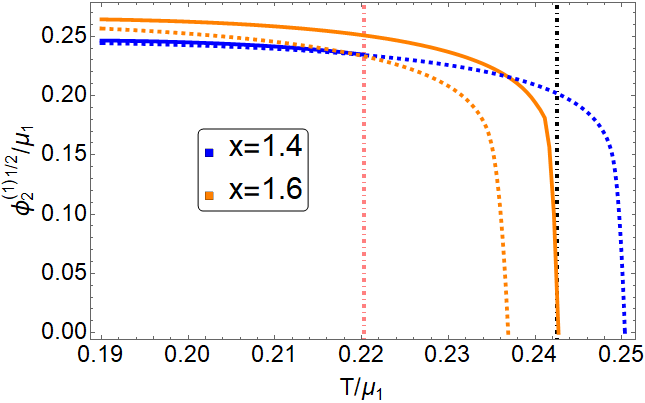}\ \hspace{0.05cm}
  \includegraphics[scale=0.31]{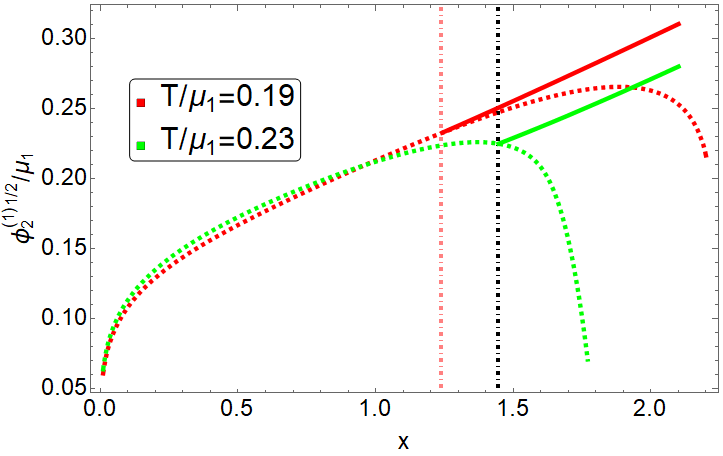}\ \hspace{0.05cm}
\caption{\label{fig13} The CDW order parameter $\phi_2^{(1)}$ as
the function of temperature $T$(left) and doping parameter
$\text{x}$(right). The solid lines in color stand for
$\phi_2^{(1)}$ in SSC phase, while the dashed lines in color stand
for $\phi_2^{(1)}$ in the absence of SC. The vertical lines
denote the location of SSC phase transition. }}
\end{figure}

\begin{figure} [h]
  \center{
  \includegraphics[scale=0.365]{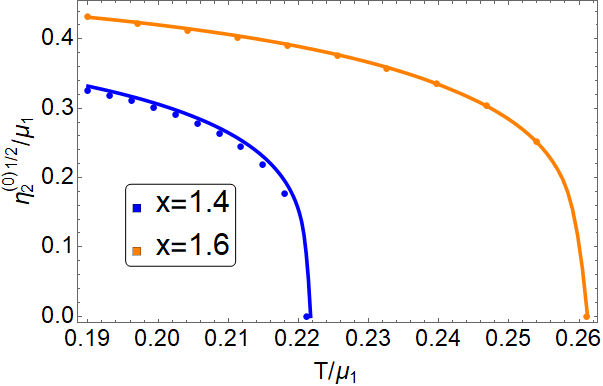}\ \hspace{0.05cm}
   \includegraphics[scale=0.376]{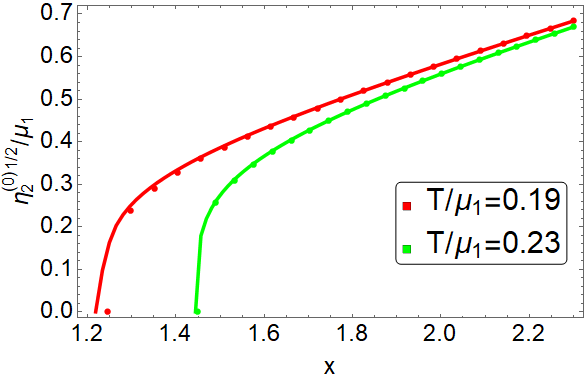}\ \hspace{0.05cm}
  \caption{\label{fig10} The uniform SC order parameter $\eta_2^{(0)}$ as the function of
temperature $T$(left) and doping parameter $\text{x}$(right).
 The solid curves in color stand for $\eta_2^{(0)}$ in SSC
phase, while the dotted curves in color stand for $\eta_2^{(0)}$
in the absence of CDW. }}
\end{figure}

\subsection{The interplay of three orders }\label{pdwcdwsc}
Next, we demonstrate the temperature behavior of the order
parameters and charge density in details, which is helpful for us
to understand the structure of the phase diagram obtained in Fig.\ref{fig6}. First of all, we are concerned with the order
parameter of PDW, namely $\eta_2^{(2)}$.   The non-zero value of
$\eta_2^{(2)}$ implies that the Cooper pairs develop a periodic
 structure with spatial oscillation due to the presence of the CDW. In
Fig.\ref{fig9}, we plot $\eta_2^{(2)}$ as the function of $T$ and
$\text{x}$, respectively. The left plot shows that below the
critical temperature the PDW grows rapidly as $T$ drops down. The
right plot shows that the PDW becomes prominent with the increase
of the doping parameter $\text{x}$ at first, and then
gradually decays with the doping parameter $\text{x}$.

\begin{figure} [h]
  \center{
  \includegraphics[scale=0.4]{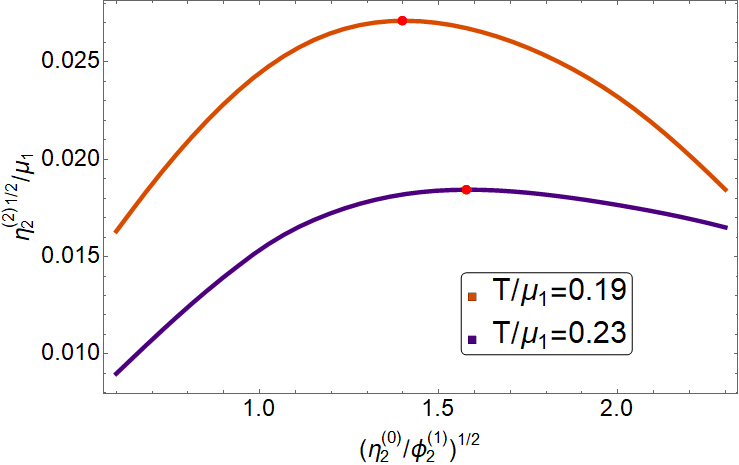}\ \hspace{0.05cm}
  \caption{\label{fig18} The relation between the PDW order parameter $\eta_2^{(2)}$ and
the ratio of uniform SC order parameter to the CDW order parameter
$\eta_2^{(0)}/\phi _{2}^{(1)}$. }}
\end{figure}

In parallel, we plot the order parameter of CDW $\phi _{2}^{(1)}$
and that of the uniform SC $\eta_2^{(0)}$ as functions of $T$ and
$\text{x}$ in Fig.\ref{fig13} and Fig.\ref{fig10}, respectively.
First of all, we are very interested in how these three
orders interact in SSC phase. In particular, we are very concerned
how the CDW order and the uniform SC order would affect the PDW
order in SSC phase. Or conversely, from the PDW order, can one
read any information about the CDW order and the uniform SC order?
As the intertwined order of CDW and uniform SC order, we know that
the PDW order $\phi _{2}^{(1)}$ must be zero when either CDW order
$\phi _{2}^{(1)}$ or uniform SC order $\eta_2^{(0)}$ vanishes. As
a result, we intend to plot the relation between $\eta_2^{(2)}$
and $\eta_2^{(0)}/\phi _{2}^{(1)}$, as illustrated in Fig.
\ref{fig18}. It is interesting to notice that there exists a tip
in each curve indeed. Moreover, with the decrease of temperature,
we find that $\eta_2^{(0)}/\phi _{2}^{(1)}$ of the tip approaches
to one, which means that the PDW order becomes prominent when the
 CDW order and uniform SC order have a balanced contribution in the
 SSC phase.

Next, we turn to investigate how the CDW order and the uniform SC
order would affect to each other in SSC phase. When the
temperature goes down, there are two possible sequences for
the occurrence of phase transition. One is that the CDW order is
formed prior to the uniform SC order, while the other is that the
uniform SC order is formed prior to the CDW order, which depends
on the doping parameter $\text{x}$. In the left plot of
Fig.\ref{fig13}, the blue curve demonstrates that the CDW order is
formed prior to the uniform SC order.
In comparison with the
data in the absence of SC, we find the saturated value of
$\phi_2^{(1)}$ grows up a little bit higher, which looks
interesting. While the orange curve demonstrates that the
formation of uniform SC order is prior to the CDW. We find that
the critical temperature of CDW in the presence of SC is higher
than that in the absence of SC, which is consistent with the phase
diagram in Fig.\ref{fig6}. Again, its saturated value becomes a
little bit higher.

On the other hand, from the left plot of Fig.\ref{fig10} we notice
that the data of the uniform SC $\eta_2^{(0)}$ in SSC phase are
the same as those in the absence of CDW. This means that the CDW
order does not affect the formation of uniform SC. The critical
temperature of SC is not sensitive to the CDW order either, as
illustrated in Fig.\ref{fig6}. This could be understood as
follows, the CDW phase is characterized by $\phi_2^{(1)}$ and
$\rho^{(2)}_{B}$. Numerically such sub-leading orders in Fourier
expansion would not change the data of leading orders much. This
also interprets why the critical temperature of CDW in SC phase is
larger than that in the absence of SC phase in Fig.\ref{fig6},
since the presence of more carriers would make it easier to
distort the crystal lattice leading to the spontaneous breaking of
translational symmetry in a doping system.

\begin{figure} [h]
  \center{
   \includegraphics[scale=0.35]{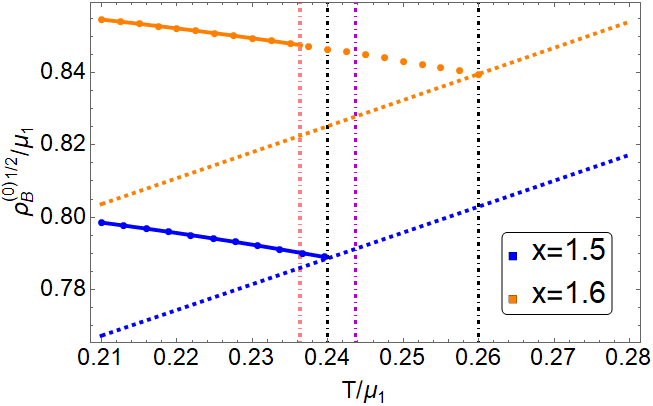}\ \hspace{0.05cm}
  \includegraphics[scale=0.32]{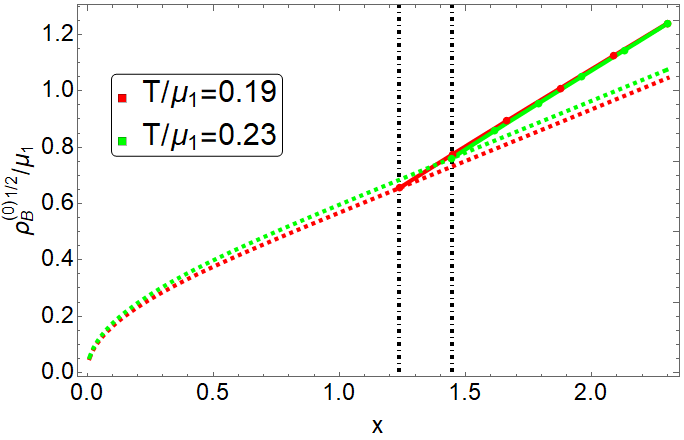}\ \hspace{0.05cm}
\caption{\label{fig14}The charge density $\rho^{(0)}_{B}$  in the
presence of CDW as the function of temperature $T$(left) and
doping parameter $\text{x}$(right). The solid curves in color
stand for $\rho^{(0)}_{B}$ in SSC phase, while the dotted curves
in color stand for $\rho^{(0)}_{B}$ in the absence of CDW, and the
dashed lines in color stand for $\rho^{(0)}_{B}$ in RN black hole.
The vertical lines in black denote the location of the critical
temperature of SC, and the vertical lines in pink or purple  denote the
location of the critical temperature of CDW.}}
\end{figure}

\begin{figure} [h]
  \center{
  \includegraphics[scale=0.32]{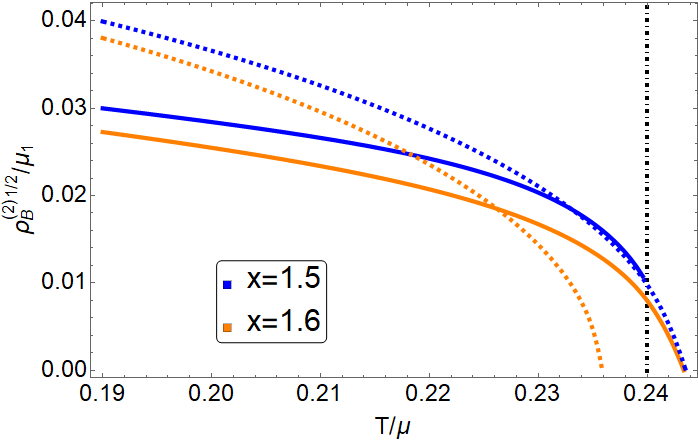}\ \hspace{0.05cm}
  \includegraphics[scale=0.33]{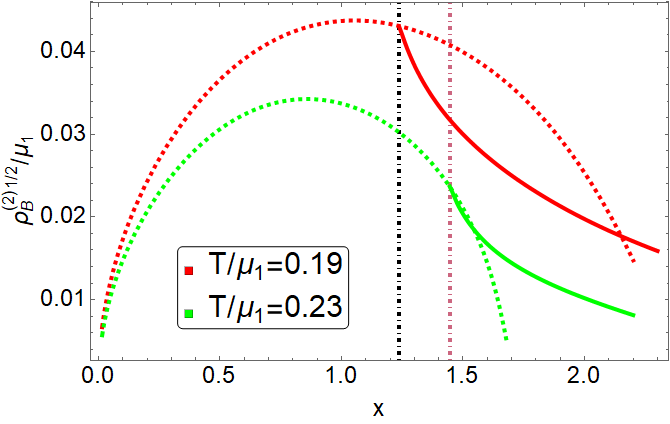}\ \hspace{0.05cm}
  \caption{\label{fig11} $\rho^{(2)}_{B}$  as the function of temperature $T$(left) and doping parameter $\text{x}$(right). The solid curves in color
stand for $\rho^{(0)}_{B}$ in SSC phase, while the dashed lines in
color stand for $\rho^{(2)}_{B}$  in the absence of SC. The
vertical lines denote the location of SSC phase transition.}}
\end{figure}

Next we plot the charge density $\rho^{(0)}_{B}$ and
$\rho^{(2)}_{B}$  as the function of temperature $T$ and
$\text{x}$. In our opinion, in SSC phase these two terms contain
the information about both the CDW order and the PDW order. In
Fig.\ref{fig14} we find $\rho^{(0)}_{B}$ is sensitive to the SC
order only, and independent of the CDW order. However, we insist
that if the formation of CDW order is prior to the SC phase, some
free charges should be pinned by the lattice effects and exhibit
an insulating behavior\footnote{Although in current paper we do
not introduce the lattice structure for the simplicity.}.
Thus, $\rho^{(0)}_{B}$ contains two ingredients in SSC phase.
One is the average charge density of CDW, and the other is the
average charge density of superfluid. In the left plot of
Fig.\ref{fig11}, the blue curve illustrates the case that the CDW
order is formed prior to the SC phase. We find $\rho^{(2)}_{B}$
jumps at the critical temperature of SC, signaling the formation
of PDW. While the orange curve illustrates the case that the
uniform SC order is formed prior to the CDW phase.  One notices
that the critical temperature of CDW in the presence of SC is
higher than that in the absence of SC, which is consistent with
the phase diagram in Fig.\ref{fig6}. In this case, PDW is formed
simultaneously with CDW. Thus from beginning, $\rho^{(2)}_{B}$
contains the information of these two orders. Obviously, the PDW
shares the same period as the CDW\footnote{This is consistent with
phenomenon observed in experiment where the period of PDW is twice
of that of CDW, because different order parameters are applied to
characterize the superconductivity. Here we take the modular part
of the complex scalar field, $\eta_2$, as the order parameter,
which is always positive definite (hence
$\eta_2^{(0)}>\eta_2^{(2)}$). However, if one takes the real part
of the complex scalar field as the order parameter (which could be
negative), for instance gauging fixing the imaginary part to zero,
then taking the Fourier expansion, one can easily find the period
of the modular is just one half of the period of real part of the
complex field.}. Furthermore, it is important to notice that the
saturated $\rho^{(2)}_{B}$ becomes smaller in comparison with that
in the absence of SC phase, which implies that the CDW order is
suppressed by the presence of SC order. In the right plot of
Fig.\ref{fig11}, we find that at low temperature $\rho^{(2)}_{B}$
decreases with the presence of the SC order as well.

\begin{figure} [h]
  \center{
  \includegraphics[scale=0.32]{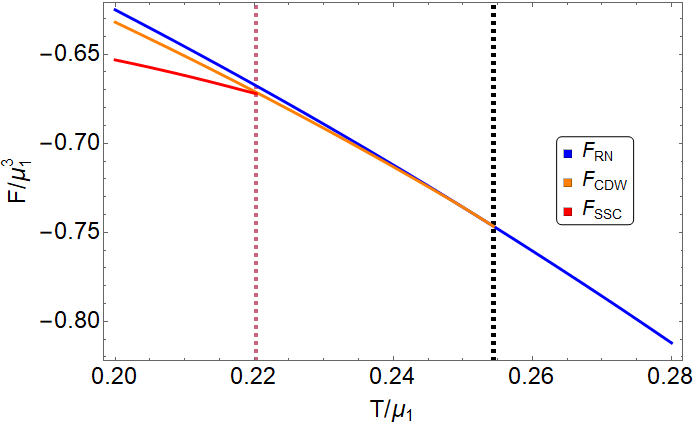}\ \hspace{0.05cm}
  \includegraphics[scale=0.4]{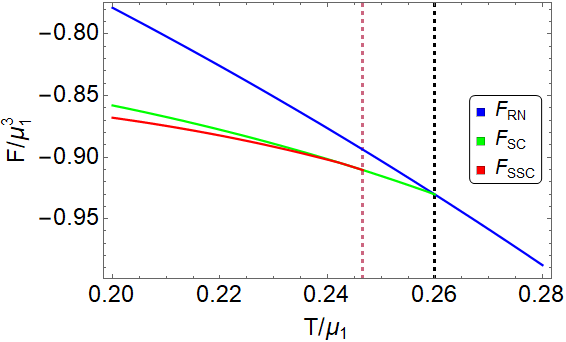}\ \hspace{0.05cm}
  \caption{\label{fig12} The averaged  free energy for RN black hole, CDW/SC  black hole and SSC black hole with $\text{x}=1.4$(left) and $\text{x}=1.6$(right), respectively.}}
\end{figure}

\subsection{The free energy of three phases}\label{pdwcdwsc}
In the end of this section we intend to verify that among above
three phases, namely CDW, SC and SSC phase, the striped
superconducting phase is the most stable state, and thus is
favored by the system. For this purpose, we compute the averaged
free energy of the system. According to
Ref.\cite{Balasubramanian:1999re,Donos:2012yu,Donos:2013wia,Withers:2013loa},
the free energy of the dual field on the boundary is identified
with the product of the temperature and the on-shell
Euclidean bulk action. Specifically, the averaged free
energy of the system is given by
\begin{equation}\label{eq:eps+3}
  \begin{aligned}
     & F=M-\mu _{1}Q_{A}-\mu _{2}Q_{B}-TS,
  \end{aligned}
\end{equation}
where
\begin{equation}\label{eq:eps+3}
  \begin{aligned}
     & M=2+\frac{\mu _{1}}{2}+\frac{\mu _{2}}{2}-\frac{3k_{c}}{2\pi }\int_{0}^{2\pi /k_{c}} q_{tt}(x)dx, \\
     & Q_{A}=\frac{k_{c}}{2\pi }\int_{0}^{2\pi /k_c} \rho _{A}(x) dx, \\
     & Q_{B}=\frac{k_{c}}{2\pi }\int_{0}^{2\pi /k_c} \rho _{B}(x) dx, \\
     &S=k\int_{0}^{2\pi /k_c} \sqrt{Q_{xx}(x,1)Q_{yy}(x,1)} dx.
  \end{aligned}
\end{equation}
In Fig.\ref{fig12}, we plot the averaged  free energy for
different background solutions with $\text{x}=1.4$ and
$\text{x}=1.6$, respectively. The system undergoes the phase
transitions from RN black hole to striped black hole with CDW and
then to SSC black hole, or from RN black hole to  black hole with
SC and then to SSC black hole. The plots evidently show that the
free energy of black hole with SSC is the lowest one in comparison
with other two phases. This indicates that the SSC branch is
thermodynamically preferred under the critical temperature,
indeed.

\section{Discussion}\label{sec3}
In this paper we have constructed a holographic model for striped
superconductor, where both $U(1)$ symmetry and translational
symmetry are spontaneously broken. Firstly, we have obtained the
phase diagram for the CDW phase and the uniform SC phase
separately, and then constructed the phase diagram for the striped
superconductor, which contains three orders, namely the CDW order,
uniform SC order and PDW order. The PDW order is implemented as
the intertwined order of the CDW order and the uniform SC order.
It is found that the CDW order is suppressed by the presence of
the uniform SC order, while the SC order is not sensitive to the
presence of CDW order. Furthermore, PDW shares the same period
with CDW. This relation is the same as that observed in
experiments \cite{Hamidian:2016}. Finally, we have also
demonstrated that among all the possible solutions, the black hole
in SSC phase has the lowest free energy and thus is favored from
the thermodynamical point of view.

It is quite instructive to compare our results obtained in this
paper with those in Ref.\cite{Ling:2019gjy}. Two different kinds
of CDWs were constructed in these two papers. In current paper CDW
contains even orders of the Fourier modes of charge density, while
in Ref.\cite{Ling:2019gjy} the CDW contains odd orders only. The
different forms of CDW lead to distinct behavior in the interplay
of CDW order and SC order. In \cite{Ling:2019gjy} due to the
absence of free charges, the $U(1)$ gauge symmetry is broken by the
presence of CDW only, thus the SC order benefits from the
existence of CDW. While in current paper the $U(1)$ gauge symmetry
is broken by the presence of free charges, and we find the uniform
SC order is not sensitive to CDW at all. We conjecture that the
CDW observed in practical materials could be the combination of
these two kinds of CDW in holographic framework and thus exhibit
more abundant interplay behavior with the SC order. Therefore, it
is quite desirable to construct a holographic model for CDW which
contains both even and odd orders of the charge density.

In current model we have only considered the essential ingredients
to generate the striped superconductor as the intertwined phase of
CDW order and SC order. It is quite worthwhile to investigate more
complicated relations between CDW and SC by introducing various
interacting terms among gauge fields and scalar fields, for
instance, considering the coupling between the condensation field
of SC and the dilaton field of CDW. This may disclose more novel
phenomena in high temperature superconductivity and deserve for
further investigation.

\section*{Acknowledgments}

We are grateful to Xian-hui Ge, Peng Liu, Yuxuan Liu, Fu-wen Shu
and Yu Tian for helpful discussions. This work is supported in
part by the Natural Science Foundation of China under Grant
No.~11875053 and 12035016.

\end{document}